\documentclass[showpacs,aps,prb,reprint,twocolumn]{revtex4-1}

\usepackage{enumitem}
\usepackage{commath}
\usepackage{amsmath,bm}
\usepackage{graphicx}
\usepackage{hyperref}
\usepackage{txfonts}

\draft 

\begin{document}


\title{Hydrodynamic effects on the energy transfer from dipoles to metal slab}
\author{Daniel Brown}
\author{Hai-Yao Deng}
\email{DengH4@cardiff.ac.uk}
\affiliation{School of Physics and Astronomy, Cardiff University, 5 The Parade, Cardiff CF24 3AA, Wales, United Kingdom}

\begin{abstract}
A systematic study of nonlocal and size effects on the energy transfer of a dipole (e.g. a molecule or a quantum dot) induced by the proximity of a metal slab is presented. Nonlocal effects are accounted for using the hydrodynamic model (HDM). We derive a general relation that connects the energy transfer rate to the linear charge density-density response function of the slab. This function is explicitly evaluated for the HDM and the local Drude model. We show that a thin metal slab can support a series of higher-frequency surface plasma wave (SPW) modes in addition to the normal SPW modes thanks to the nonlocal effects. These modes markedly alter the response and the energy transfer process, as revealed in the structure of the energy transfer rate in the parameter space. Our findings are important for applications such as the recently developed metal-induced energy transfer imaging, which relies on accurate modeling of the energy transfer rate.  
\end{abstract}

\maketitle

\section{Introduction}
\label{sec:1}
As well known since the 1970s, the radiation properties of a dipole, which may be represented by a molecule or a quantum dot (QD), can be strongly affected by a nearby metal~\cite{morawitz1969,kuhn1970,barton1970,chance1974,morzwitz1974,chance1975,philpott1975,moskovits1978}. For example~\cite{pineda1985,nustovit2012}, fluorescence can be quenched as the dipole channels most of its energy to the metal in the form of charge density waves propagating along the metal surface, known as surface plasma waves (SPW). These waves can also modify the dipole-dipole interaction and enable coherent energy transmission between dipoles far apart~\cite{philpott1975,stuart1998}. Using conducting structures to obtain desired properties of an adjacent object has become an versatile tool in a plethora of research areas~\cite{kimble2008,curto2010,dey2018,grob2018,chandra2020,lu2020}. 

In the past few years, a powerful sub-nanometer imaging technique, called Metal-Induced Energy Transfer Imaging (MIETI) was developed~\cite{chizhik2014,karedla2014,chizhik2017,baronsky2017,isbaner2018,ghosh2019}, utilizing the sensitive dependence of the fluorescence time (rather than the radiative power) on the dipole-metal distance. MIETI relies on accurate modeling of the dipole-to-metal energy transfer rate. However, most theoretical studies~\cite{morawitz1969,kuhn1970,barton1970,chance1974,morzwitz1974,chance1975,philpott1975,moskovits1978,cao2021} treat the metal as a simple dielectric by essentially the Drude model and non-local effects have been ignored. In studies where those effects were included~\cite{vagov2016,Deng2017a}, only a semi-infinite metal was considered rather than a metal slab, which is usually the geometry in reality~\cite{wu2009}. 

In the present work, we systematically explore the non-local effects, by means of a hydrodynamic model (HDM)~\cite{barton1979,pendry2013,Deng2019,Deng2020}, on the energy transfer rate as a function of the dipole-metal distance and the slab thickness. We find that hydrodynamic effects -- together with size effects -- can significantly impact the energy transfer process, entailing a very different picture than ignoring the effects. In this investigation, we also provide a general relation between the transfer rate and the electrodynamic response function of the metal, which is applicable beyond the HDM. 

In the next section, we describe the system and the formalism, explaining the connection between the energy transfer rate and the charge density-density response function of the metal and also giving explicit expressions for the energy transfer rate in the HDM and the LDM. In Sec.~\ref{sec:3}, we present comprehensive numerical results, revealing the existence of a sequence of higher-frequency SPW branches and demonstrating their significance in the structure of the energy transfer rate in the parameter space spanned by the intrinsic dipole frequency, the slab thickness and the dipole-slab distance. We summarize the paper in Sec.~\ref{sec:4}, wherein we also discuss experimental relevance of the results. 

\section{System and formalism}
\label{sec:2}
\begin{figure}
\begin{center}
\includegraphics[width=0.48\textwidth]{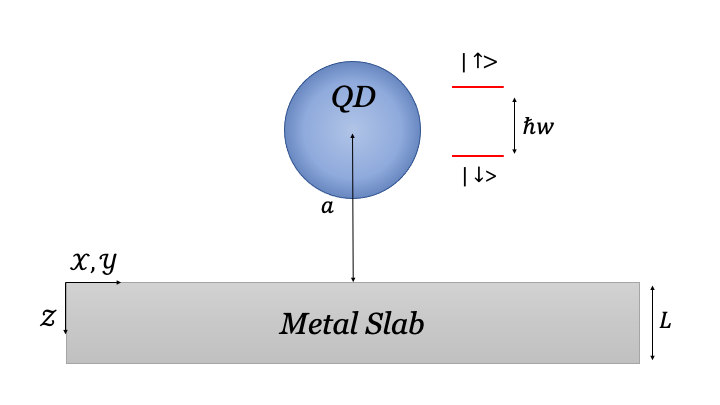}
\end{center}
\caption{Schematic of the system: a dipole, represented by a QD is placed a distance $a$ over a metal slab of thickness $L$. The QD has two energy levels $|\uparrow\rangle$ and $\downarrow\rangle$, apart by $\hbar\omega$ in energy. The two surfaces of the slab are located at $z=0$ and $z=L>0$, respectively. \label{fig:0}}
\end{figure}

The system under study is depicted in Fig.~\ref{fig:0}, where a dipole exemplified by a two-level QD is placed at a distance $a$ over the surface of a metal slab of thickness $L$. The two levels of the QD are denoted by $|\uparrow\rangle$ and $|\downarrow\rangle$, respectively, which are separated in energy by $\hbar\omega$, where $\hbar$ is the reduced Planck constant and $\Omega$ is the associated frequency. The Hamiltonian of the system can be written as a sum of three parts, $H_{QD} = \frac{\hbar\omega}{2}\left(|\uparrow\rangle\langle\uparrow| - |\downarrow\rangle\langle\downarrow|\right)$, $H_\text{Metal} = \sum_M \varepsilon_M|M\rangle\langle M|$ and $H'$, which describe the QD, the metal and the coupling between them, respectively. Here $M$ labels the eigenstates, $|M\rangle$, and eigenvalues, $\varepsilon_M$ of $H_\text{Metal}$. We shall assume that the distance $a$ is not too small so that electronic tunneling can be ignored between the QD and the metal. We also neglect retardation effects. As such, the coupling is of electrostatic character and can be written as 
\begin{equation}
H' = e\int d^3\mathbf{x} d^3\mathbf{y} \frac{\hat{\rho}(\mathbf{x}) \hat{n}(\mathbf{y})}{\abs{\mathbf{x}-\mathbf{y}}},
\end{equation} 
where $e$ denotes the charge of an electron, $\hat{n}$ is the electron density operator subtracted by the mean density of electrons in the QD and $\hat{\rho}$ is the charge density operator for the metal. 

Now we imagine the QD being excited to its upper state $|\uparrow\rangle$ by absorbing a photon and set to calculate the rate at which the absorbed energy is transferred to the metal by the coupling Hamiltonian $H'$. We denote the rate by $\frac{1}{\tau}$, where $\tau$ may be called the non-radiative lifetime of the QD. The fluorescence life time can be calculated as $\frac{\tau \tau'}{\tau + \tau'}$, where $\tau'$ denotes the radiative lifetime, which is to be treated in a future publication. In the rest of this section, we describe the formalism for calculating $\tau$ as a function of the system parameters. 

\subsection{Connection with the density-density response function}
\label{sec:2.1}
The initial state for the global system is written as $|\uparrow,M\rangle = |\uparrow\rangle \otimes |M\rangle$, where the metal has been assumed in state $|M\rangle$. Under the action of $H'$, the QD can jump to the lower state $|\downarrow\rangle$ by giving out energy to the metal which synchronously transits to another state $|M'\rangle$. The transition amplitude reads
\begin{eqnarray}
\langle\downarrow,M'|H'|\uparrow,M\rangle = \int d\mathbf{x} ~ \Phi^*(\mathbf{x}) \hat{\rho}_{M',M}(\mathbf{x}),
\end{eqnarray}
where $\hat{\rho}_{M',M}(\mathbf{x}) = \langle M'|\hat{\rho}(\mathbf{x})|M\rangle$ and
\begin{equation}
\Phi(\mathbf{x}) = e\int d\mathbf{y} \frac{\langle \uparrow|\hat{n}(\mathbf{y})|\downarrow \rangle}{|\mathbf{y} - \mathbf{x}|}
\end{equation}
is the potential generated by the QD. The total rate that the QD transfers its energy to the metal can be obtained from Fermi's golden rule, which gives
\begin{eqnarray}
\frac{1}{\tau} = \frac{2\pi}{\hbar} \sum_{M,M'}p_M \abs{\int d\mathbf{x} ~ \Phi(\mathbf{x}) \hat{\rho}_{M,M'}(\mathbf{x})}^2 \delta(\hbar\omega + \epsilon_{M} - \epsilon_{M'}). \label{4}
\end{eqnarray}
Here we have assumed that the metal is initially in thermodynamic equilibrium characterized by the Boltzmann factor $p_M = e^{-\varepsilon_M/k_BT}/Z$ with $Z$ being the partition function, $k_B$ the Boltzmann constant and $T$ the temperature, and $\delta(x)$ denotes the Dirac function. 

We can relate $\frac{1}{\tau}$ as given in Eq.~(\ref{4}) to the charge density-density response function, denoted by $\chi_\omega(\mathbf{x},\mathbf{y})$ of the metal. Physically, $\chi_\omega(\mathbf{x},\mathbf{y})$ gives the charge density at $\mathbf{x}$ induced by an \textit{external} oscillatory (at frequency $\omega$) electrostatic potential localized at $\mathbf{y}$. Quantum linear response theory gives~\cite{Deng2020,lubensky}
\begin{small}
\begin{equation}
\chi_\omega(\mathbf{x},\mathbf{y}) = \sum_{M,M'} p_M \left(\frac{\hat{\rho}_{M,M'}(\mathbf{x})\hat{\rho}_{M',M}(\mathbf{y})}{\hbar\omega + i0_+ + \epsilon_M - \epsilon_{M'}} - \frac{\hat{\rho}_{M,M'}(\mathbf{y})\hat{\rho}_{M',M}(\mathbf{x})}{\hbar\omega + i0_+ + \epsilon_{M'} - \epsilon_{M}}\right). 
\end{equation}
\end{small}
Using Im$\left(\frac{1}{x+i0_+}\right) = -\pi \delta(x)$ and the fluctuation-dissipation theorem~\cite{lubensky}, one may readily verify that 
\begin{eqnarray}
\frac{1}{\tau} = \frac{2}{\hbar} \frac{1}{e^{-\beta\hbar\omega} - 1} \mbox{Im}\left(\Gamma_\omega\right) \approx - \frac{2}{\hbar}~ \mbox{Im}\left(\Gamma_\omega\right), \label{6}
\end{eqnarray}
where the approximation occurs at temperatures much lower than $\hbar\omega/k_B$, which is assumed throughout the paper, and
\begin{equation}
\Gamma_\omega = \int d\mathbf{x} \int d\mathbf{y} ~ \Phi(\mathbf{x}) \chi_\omega(\mathbf{x},\mathbf{y}) \Phi^*(\mathbf{y}). \label{7}
\end{equation}
Equations (\ref{6}) and (\ref{7}) are generally valid, not just for the QD coupled to a metal but to any material, as long as the appropriate $\chi_\omega$ is supplied. Equation (\ref{7}) describes the net effect of the charges induced at point $\mathbf{y}$ by the QD propagating to $\mathbf{x}$ whereby they interact with the QD again. 

\begin{figure*}
\begin{center}
\includegraphics[width=0.95\textwidth]{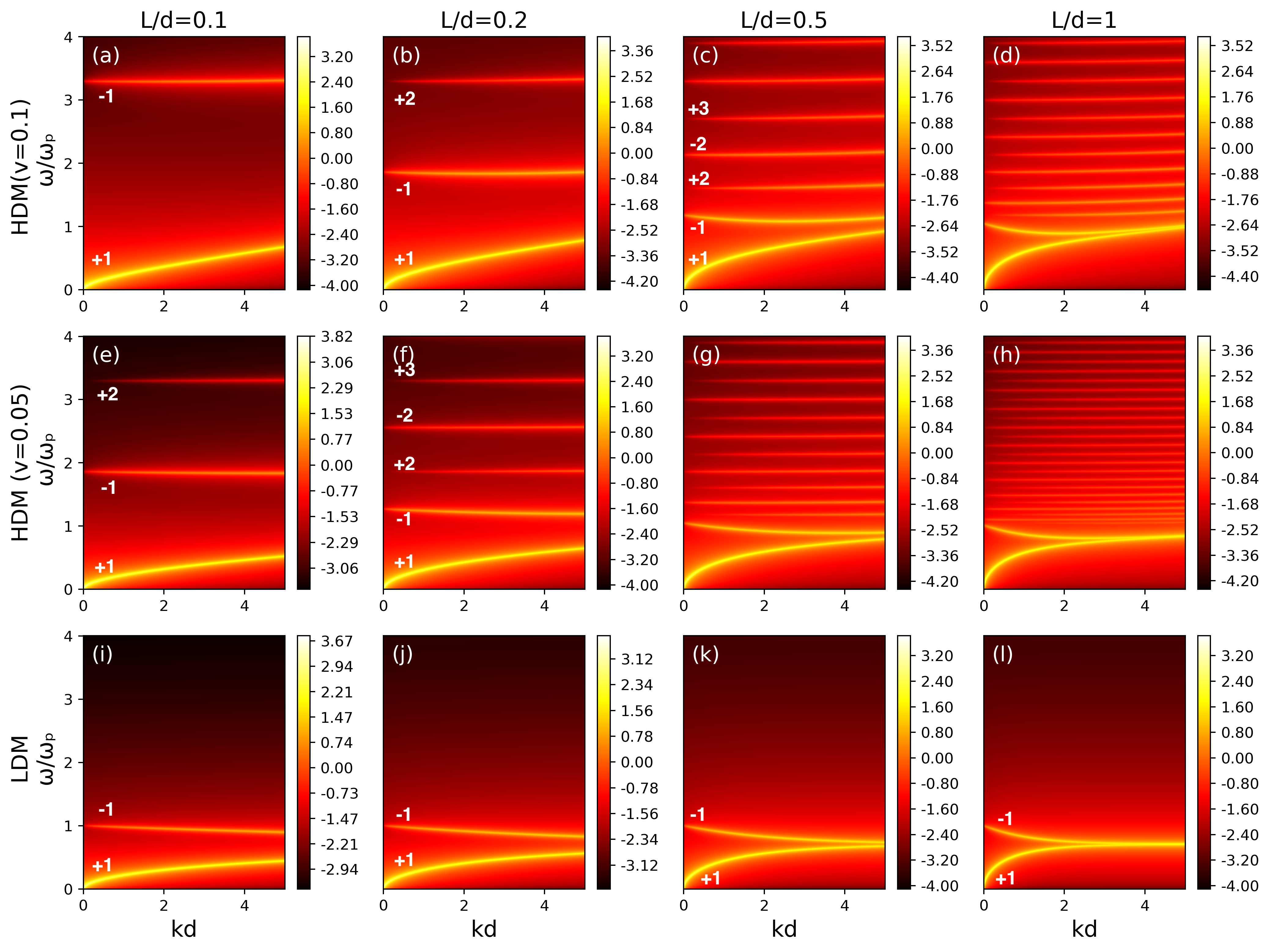}
\end{center}
\caption{Map of $\log_{10}\left[\mbox{Im}\left(\frac{1}{\epsilon_+(k,\omega)}\right) + \mbox{Im}\left(\frac{1}{\epsilon_-(k,\omega)}\right)\right]$, which reveals the dispersion of SPWs admitted on a slab of thickness $L$. SPW branches are indicated by the numbers. The non-locality parameter $v$ is expressed in unit of $\omega_pd$. Very small damping $\gamma/\omega_p = 0.01$ is used. Unlike LDM, HDM allows higher-frequency modes to appear on a slab. \label{fig:1}}
\end{figure*} 

\subsection{The dipole approximation}
\label{sec:2.2}
Considering that the QD is small, we may invoke the dipole approximation for $\Phi(\mathbf{x})$, i.e.
\begin{equation}
\Phi(\mathbf{x}) \approx \mathbf{D}\cdot \partial_{\mathbf{x}_0} \frac{1}{\abs{\mathbf{x}_0 - \mathbf{x}}}, \label{dip}
\end{equation}
where $\mathbf{x}_0 = (\mathbf{r}_0,z_0)$ with $\mathbf{r}_0 = (0,0)$ and $z_0=-a<0$ locates the center of the QD, and $\mathbf{D} = (D_x,D_y,D_z)$ denotes the dipole transition element of the QD, given by
\begin{equation}
\mathbf{D} = e \int d\mathbf{y} ~ \mathbf{y} \langle \uparrow|\hat{n}(\mathbf{y})|\downarrow\rangle, 
\end{equation}
Note that the integral here is virtually carried over the volume of the QD. 

We may make use of the translation symmetry along the slab surfaces by writing   
\begin{equation}
\chi_\omega(\mathbf{x},\mathbf{y}) = \int \frac{d^2\mathbf{k}}{4\pi^2}~\chi_{\mathbf{k}\omega}(z,z') e^{i\mathbf{k}\cdot(\mathbf{r}-\mathbf{r}')}. \label{ftk}
\end{equation}
Here $\mathbf{x} = (\mathbf{r},z)$, $\mathbf{y} = (\mathbf{r}',z')$ and $\mathbf{k}$ is a wave vector. Also we introduce a double cosine Fourier transform, defined as
\begin{equation}
\chi_{\mathbf{k}\omega}(z,z') = \sum^\infty_{m,n = 0} \frac{1}{L_m} \chi_{\mathbf{k}\omega}(n,m) \cos(q_nz) \cos(q_mz'), \label{ftq}
\end{equation}
Here $q_n = \frac{n\pi}{L}$ and $L_m = \frac{L}{2-\delta_{m,0}}$ with $\delta_{m,m}$ being the Kronecker symbol. Physically, $e^{i(\mathbf{k}\cdot\mathbf{r}-\omega t)}\chi_{\mathbf{k}\omega}(n,m)$ gives the charge density induced by an \textit{external} electrostatic potential of this form: $e^{i(\mathbf{k}\cdot\mathbf{r}-\omega t)}\cos(q_mz)$. Via Eqs.~(\ref{dip}), (\ref{ftk}) and (\ref{ftq}), one arrives at
\begin{widetext}
\begin{equation}
\Gamma_\omega = \sum_{m,n}\sum_{\mu,\nu}D_\mu D^*_\nu \int d^2\mathbf{k} \frac{\chi_{\mathbf{k}\omega}(n,m)}{L_m} \frac{\bar{k}_\mu e^{-ka}}{k^2+q^2_n} \frac{\bar{k}^*_\nu e^{-ka}}{k^2+q^2_m} \left(1-(-1)^ne^{-kL}\right)\left(1-(-1)^me^{-kL}\right),
\end{equation}
\end{widetext}
where $k = |\mathbf{k}|$ and $\bar{\mathbf{k}} = (i\mathbf{k},k)$. Note that the factor $e^{-ka}$ puts an effective cutoff on $k$: Fourier components with $k>1/a$ cannot make significant contributions. 

To make progress, we assume that the slab is invariant under reflection about its mid-plane at $z=L/2$. It follows that quantities with different parities cannot be mixed~\cite{Deng2017b}. In particular, $\chi_{\mathbf{k}\omega}(n,m)$ must vanish if $m$ and $n$ have different parities. It suggests to split $\chi_\omega$ into two sectors, the symmetric one $\chi^+_{\mathbf{k}\omega}(l,l') = \chi_{\mathbf{k}\omega}(2l,2l')$ and the anti-symmetric one $\chi^-_{\mathbf{k}\omega}(l,l') = \chi_{\mathbf{k}\omega}(2l+1,2l'+1)$, where $l,l'=0,1,...$. Additionally, we introduce the following quantities,
\begin{equation}
P^s_{k\omega}(l) = \sum^\infty_{l'=0} \frac{\chi^s_{\mathbf{k}\omega}(l',l)}{k^2 + (q^s_{l'})^2}, ~ P_{s,\omega}(k) = \sum_l \frac{4\pi}{L^s_l}\frac{P^s_{k\omega}(l)}{k^2+(q^s_l)^2}, \label{P}
\end{equation}
where $s=\pm$ and $q^+_l = q_{2l}$ and $q^-_l = q_{2l+1}$. We have considered the fact that $\chi^s_{\mathbf{k}\omega}(l)$ does not depend on the direction of $\mathbf{k}$ due to rotational symmetry about the $z$-axis. These symmetry considerations allow us to finally get 
\begin{equation}
\frac{1}{\tau} = \frac{1}{2}\frac{e^2}{\hbar d} \left(1+\cos^2\theta\right) R(\omega,a,L), \label{rat}
\end{equation}
where $\cos\theta = D_z/\abs{\mathbf{D}}$ and $d = \abs{\mathbf{D}}/e$ is a length comparable to the dimension of the QD, and 
\begin{equation}
R(\omega,a,L) = -\sum_s \int^\infty_0 dk ~ k^3 e^{-2ka} \left(1-se^{-kL}\right)^2 ~ \mbox{Im}\left(P_{s,\omega}(k)\right) \label{R}
\end{equation}
is a dimensionless quantity. In obtaining the expression for $R$, we have performed an integration over the angle of $\mathbf{k}$ and, in each quantity in the integrand the unit of length has been set with $d=1$ to simplify the notation. 

A few remarks are in order. Equation~(\ref{R}) shows that the QD probes the quantity $P_{s,\omega}(k)$ rather than $\chi_{\mathbf{k}\omega}$. This accords with the general observation~\cite{Deng2020} that entities (e.g. the QD here) residing outside a metal can only probe $P^s_{k\omega}$. Secondly, the frequency, $\frac{e^2}{\hbar d}$, should be comparable to the QD intrinsic frequency $\omega$, as the latter is roughly set by the size of the QD, which is of the same order as $d$. Finally, Eq.~(\ref{rat}) shows that the orientation of $\mathbf{D}$ also affects the energy transfer rate, but only by a factor of $2$ at the most. 

\begin{figure*}
\begin{center}
\includegraphics[width=0.95\textwidth]{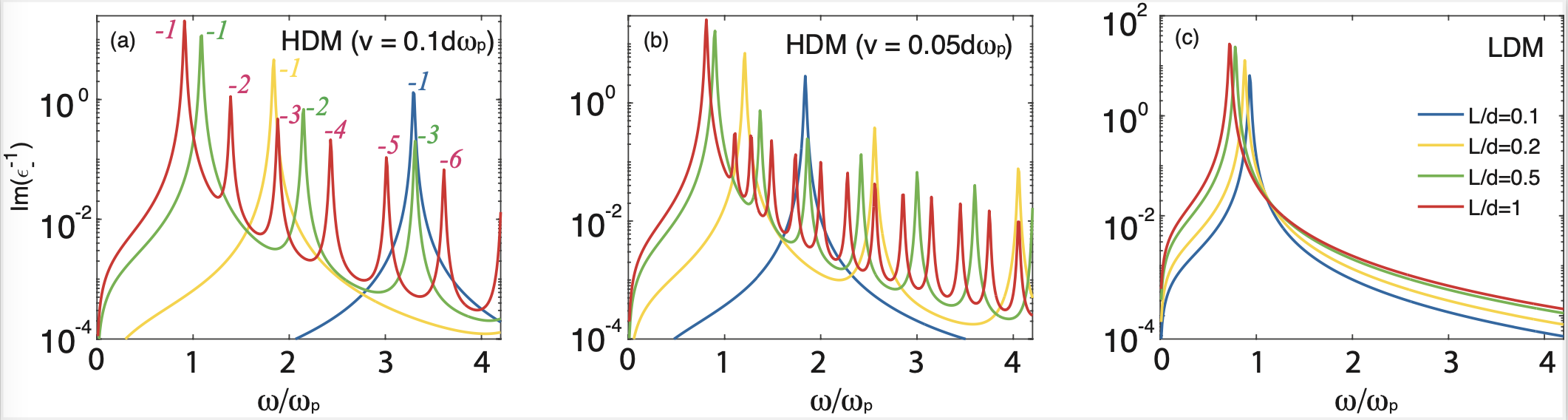}
\end{center}
\caption{Anti-symmetric modes manifested with Im$\left({\frac{1}{\epsilon_{-}(k,\omega)}}\right)$ at fixed $kd=3$ for various slab thickness $L$ and non-locality parameter $v$. Numbers in (a) indicate SPW branches. $\gamma/\omega_p = 0.01$. Legends for (a) and (b) are the same as in (c). \label{fig:2}}
\end{figure*} 

\subsection{Hydrodynamic model (HDM)}
\label{sec:2.3}
HDM treats the electrons in a metal as a fluid, governed by the linearized Navier-Stokes equation~\cite{barton1979,pendry2013,Deng2019,Deng2020}
\begin{equation}
mn_0\left(\partial_t + \gamma\right)\mathbf{V} = n_0e\mathbf{E} - mv^2\nabla n. \label{HDM} 
\end{equation}
Here $m$ is the mass of an electron, $n_0$ the mean electron density, $n$ denotes the deviation from $n_0$, $\gamma$ is a phenomenological parameter accounting for dissipation (viscous effects), $\mathbf{E}(\mathbf{x},t)$ the total electric field felt by the electrons and $\mathbf{V}(\mathbf{x},t)$ their velocity field. In addition, $v$ is a parameter that controls non-local effects. The local Drude model (LDM) is recovered as a limit of the HDM with $v=0$. 

In the Appendix, we show that for HDM
\begin{eqnarray}
P^s_{k\omega}(l) = \frac{\omega^2_p}{4\pi} \frac{1}{\tilde{\omega}^2 - \omega^2_p - v^2(q^s_l)^2} \frac{1}{\epsilon_s(k,\omega)}, \label{17}
\end{eqnarray}
where $\omega_p = \sqrt{4\pi n_0 e^2/m}$ denotes the characteristic frequency of the metal, and
\begin{equation}
\epsilon_s(k,\omega) = \frac{\bar{\omega}^2 - \Omega^2_s}{\bar{\omega}^2 - \omega^2_p}, \quad \Omega_s = \omega_s \sqrt{1+\frac{k}{Q}\frac{1+s e^{-QL}}{1-s e^{-QL}}}. \label{eps}
\end{equation}
Here $\omega_s = \frac{\omega_p}{\sqrt{2}}\left(1-se^{-kL}\right)$ and $ Q = v^{-1}\sqrt{\omega^2_p - \tilde{\omega}^2}$ with $\tilde{\omega}^2 = \bar{\omega}^2 - k^2v^2$ and $\bar{\omega} = \omega + i\gamma$. Using the identity that 
\begin{equation}
\sum^\infty_{l=0} \frac{1}{L^s_l} \frac{1}{q^2 + (q^s_l)^2} = \frac{1}{2q} \frac{1 + s e^{-qL}}{1 - s e^{-qL}}, \label{id}
\end{equation}
which is an even function of $q$, one easily finds 
\begin{equation}
P_{s,\omega}(k) = \left(1-\frac{1}{\epsilon_s(k,\omega)}\right)\frac{1}{k\left(1-se^{-kL}\right)}. 
\end{equation}
As such, we get
\begin{equation}
R(\omega,a,L) = \sum_s \int^\infty_0 dk ~ k^2 e^{-2ka} \left(1 - se^{-kL}\right) ~\mbox{Im}\left(\frac{1}{\epsilon_s(k,\omega)}\right). \label{HDMR}
\end{equation}
This equation, together with Eqs.~(\ref{rat}) and (\ref{eps}), completely determines the energy transfer rate $1/\tau$. 

Equation (\ref{HDMR}) implies that the elementary excitations that can be detected by the QD are all ciphered in the quantity $\epsilon_s(k,\omega)$, the zeros of which in the complex $\omega$-plane locate the frequency and damping rate of SPWs propagating with wavenumber $k$ along the slab surfaces, satisfying $\bar{\omega} = \Omega_s(\bar{\omega})$. 

The results for LDM are particularly simple: With $v=0$, $\Omega_s$ reduces to $\omega_s$ and then 
\begin{equation}
\epsilon_s = \frac{\bar{\omega}^2-\omega^2_s}{\bar{\omega}^2 - \omega^2_p}, \quad P^s_{k\omega}(l) = \frac{\omega^2_p}{4\pi} \frac{1}{\bar{\omega}^2-\omega^2_s}. \label{ldm}
\end{equation}
The zeros of $\epsilon_s$ occur at $\omega = \omega_s - i\gamma$. Thus, $\omega_s$ and $\gamma$ give the reputed frequency and the damping rate, respectively, of SPWs admitted on a slab by LDM. For $k\gg L^{-1}$, $\omega_s(k)$ reduces to $\omega_p/\sqrt{2}$ and becomes dispersion-less. Splitting the integral in Eq.~$(\ref{HDMR})$ into two parts, one over $k\in[0,1/L]$ and the other $k\in[L^{-1},\infty]$, one may see that the latter part gives a contribution to $R$ going as $\frac{1}{2}\mbox{Im}\left(\frac{\omega^2_p}{\omega^2_p - 2\bar{\omega}^2}\right)\left(\frac{d}{a}\right)^3$, which blows up to infinity for $a$ tending to zero regardless of whether $\omega$ is in resonance or not. Divergence of this kind is well known~\cite{horsley2014} and removed by nonlocal effects in HDM.  

\section{Numerical results and discussions}
\label{sec:3}

\subsection{Properties of surface plasma waves}
\label{sec:3.1}

The QD is localized in space and hence can give out energy to excite SPWs of any wavenumber. As is clear from Eq.~(\ref{HDMR}), the efficiency for SPWs of wavenumber $k$ to get excited by the QD is largely set by the spectral weight
\begin{equation}
\mbox{Im}\left(\frac{1}{\epsilon_s(k,\omega)}\right) = \frac{\epsilon''_s(k,\omega)}{\left[\epsilon'_s(k,\omega)\right]^2 + \left[\epsilon''_s(k,\omega)\right]^2}, \label{dis}
\end{equation}
where we have put $\epsilon_s = \epsilon'_s - i\epsilon''_s$ with $\epsilon'_s$ and $-\epsilon''_s$ being the real and imaginary parts of $\epsilon_s$, respectively. The imaginary part $\epsilon''_s$ originates from dissipation of SPWs into heat, which is ultimately traced back to the damping parameter $\gamma$ in HDM. As $\gamma$ is typically small, $\epsilon''_s$ is also generally small. Equation~(\ref{dis}) then displays a resonance wherever $\epsilon'_s(k,\omega) = 0$, which determines the SPW dispersion relation. 

To reveal the SPW dispersion, we have produced a map of dB~$= \log_{10}\left[\mbox{Im}\left(\frac{1}{\epsilon_+(k,\omega)}\right) + \mbox{Im}\left(\frac{1}{\epsilon_-(k,\omega)}\right)\right]$ in the $k-\omega$ plane, as exhibited in Fig.~\ref{fig:1}, for various values of the non-locality parameter $v$. The SPW resonances coalescence into bright lines of high dB, representing different branches. 

It is usually thought that there are merely two branches of SPWs living on a slab, stemming from the symmetric ($s=+$) and the anti-symmetric ($s=-$) sectors, respectively. This is obviously true with LDM, corresponding to $v=0$, for which the SPW has dispersion given by $\omega_s = \frac{\omega_p}{\sqrt{2}}\sqrt{1-se^{-kL}}$ as already mentioned in the preceding section [cf.~Eq.(\ref{ldm})]. For small $kL$, $\omega_+ \propto \sqrt{kL}$ whereas $\omega_- \approx \omega_p$. For large $kL$, both $\omega_+$ and $\omega_-$ tend to $\omega_p$. These features are well borne out in the bottom row of panels (i) - (l) in Fig.~\ref{fig:1}. For convenience, we shall label these ordinary branches by $l = +1$ and $l = -1$, respectively, see the figure.  

The SPW properties by HDM are strikingly different. In addition to the normal branches seen in LDM ($l = \pm 1$), there are higher-frequency branches, which we label as $l = \pm2, \pm3, ...$ in the order of increasing frequency (see Fig.~\ref{fig:1}). The branches with $l>0$ all derive from the symmetric sector $\epsilon_+$ while those with $l<0$ from the anti-symmetric sector $\epsilon_-$. They have less intensity than the primary branches $l=\pm 1$ but can still be clearly seen in the top and middle rows of panels (a) - (h) in Fig.~\ref{fig:1}. In Fig.~\ref{fig:2}, we plot as an illustration $\mbox{Im}\left(\frac{1}{\epsilon_{-}(k,\omega)}\right)$ at fixed $kd=3$ versus $\omega$ to show that the branches $l<-1$ are absent from LDM. Similar plot can be generated for the symmetric branches $l>1$, which are also absent from LDM. Thus, the modes $|l|>1$ arise from hydrodynamic effects. 

The symmetric and anti-symmetric branches are seen alternating with one another, and they are rather evenly spaced in frequency. The spacing decreases with increasing slab thickness $L$. Upon decreasing $L$, their frequencies all (including that of mode $l=+1$) shift upward dramatically. Actually, for very small $L$, even the lowest anti-symmetric branch $l=-1$ can be pushed way higher than $\omega_-$ (see the first column of panels, (a), (e) and (i) in Fig.~\ref{fig:1}). On the other hand, for large $L$, these higher-frequency branches get closer and less pronounced and could be easily erased by damping. We note in passing that the symmetric branches have less intensity (dB) at long wavelengths (i.e. small $k$) than the anti-symmetric ones, a feature that allows us to tell them apart by simple inspection.  

\begin{figure}
\begin{center}
\includegraphics[width=0.45\textwidth]{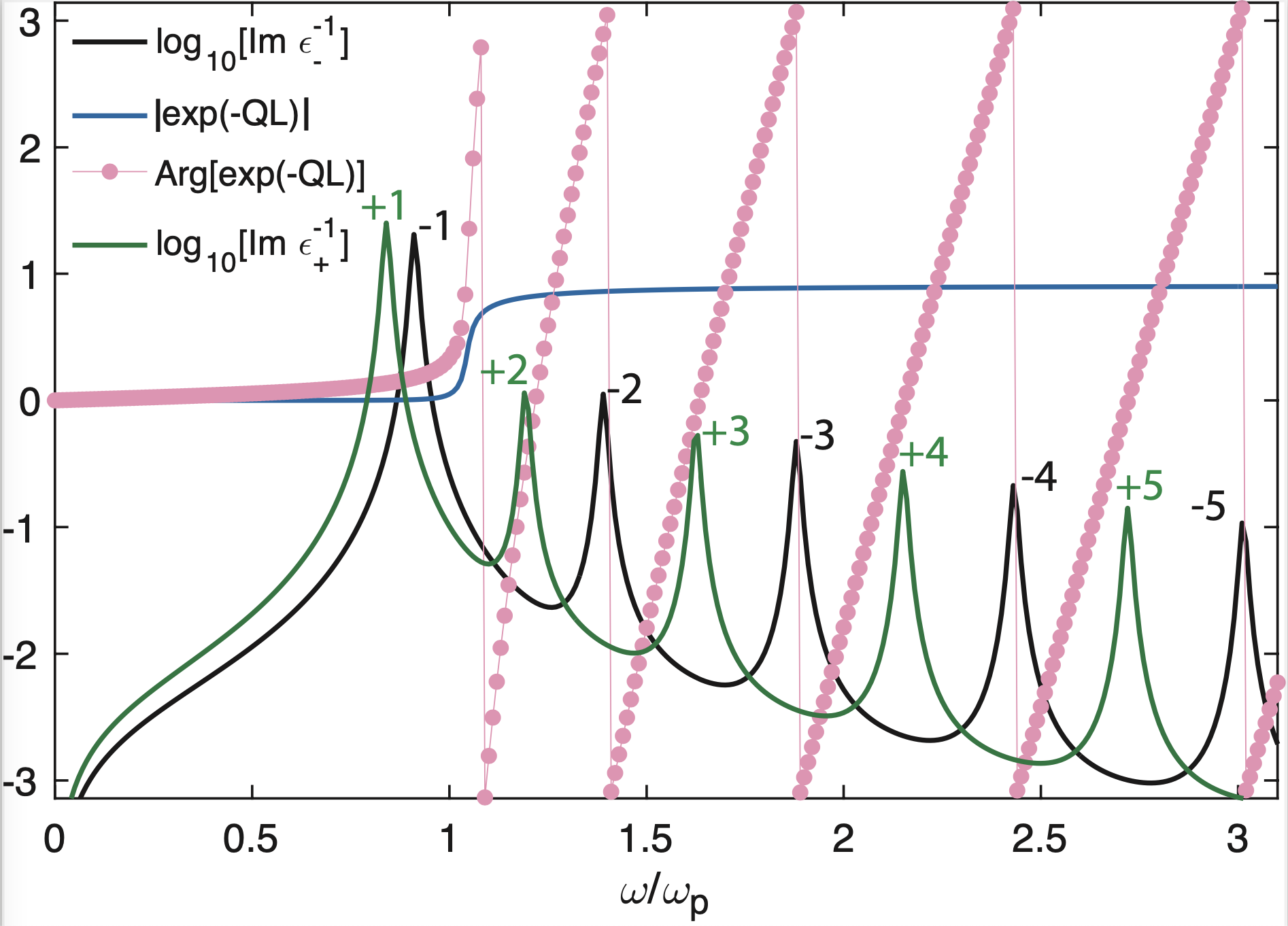}
\end{center}
\caption{Higher-frequency SPW modes appear wherever the factor $\exp(-QL)$ changes its phase from $\pi$ to $-\pi$, which gives rise to the anti-symmetric modes $l<0$, or wherever its phase angle is zero, which gives rise to the symmetric modes $l>1$. Numbers indicate branches. Parameters: $v=0.1d\omega_p$, $L/d=1$, $\gamma = 0.01\omega_p$ and $kd=3$. \label{fig:2e}}
\end{figure} 

Thanks to the hydrodynamic effects, the dispersion of the branch $l=-1$ is not monotonic in HDM [Fig.~\ref{fig:1} (d)]: the frequency decreases first and then increases with increasing $k$. The branch $l=+1$ can also undergo a change in its dispersion if $v$ is big, which is almost linear for small $L$ [Fig.~\ref{fig:1} (a)] but becomes concave as $L$ increases [Fig.~\ref{fig:1} (b) - (d)].     

Properties of the higher-frequency branches can be understood from the behaviors of the factor $\exp(-QL)$ in Eq.~(\ref{eps}). As demonstrated in Fig.~\ref{fig:2e}, the branches $l>1$ occur wherever the phase angle of $\exp(-QL)$ approaches zero, while the modes $l<-1$ occur wherever the phase angle crosses $\pi$. As a result, the symmetric and anti-symmetric branches alternate in frequency. The phase difference by $\pi$ comes from the minus sign picked up when going from $s=+1$ to $s=-1$, as is clear from Eq.~(\ref{eps}). For $\omega>\omega_p$, $kv\ll \omega$ and $\gamma/\omega_p\ll1$, we find $Q \approx \frac{i\omega_p}{v}\sqrt{(\omega/\omega_p)^2 -1}$ having neglected a small imaginary part $\sim \gamma/v$. Under such circumstance $\exp(-QL)$ is purely oscillatory with $\omega$. For $\omega/\omega_p\gg1$, the oscillation period approximates a constant $\Delta\omega = 2\pi v/L$, which explains why these branches are nearly evenly spaced. Meanwhile, it shows that the spacing decreases with decreasing $v/L$, as observed in Figs.~\ref{fig:1} and \ref{fig:2}. 

\begin{figure}
\begin{center}
\includegraphics[width=0.45\textwidth]{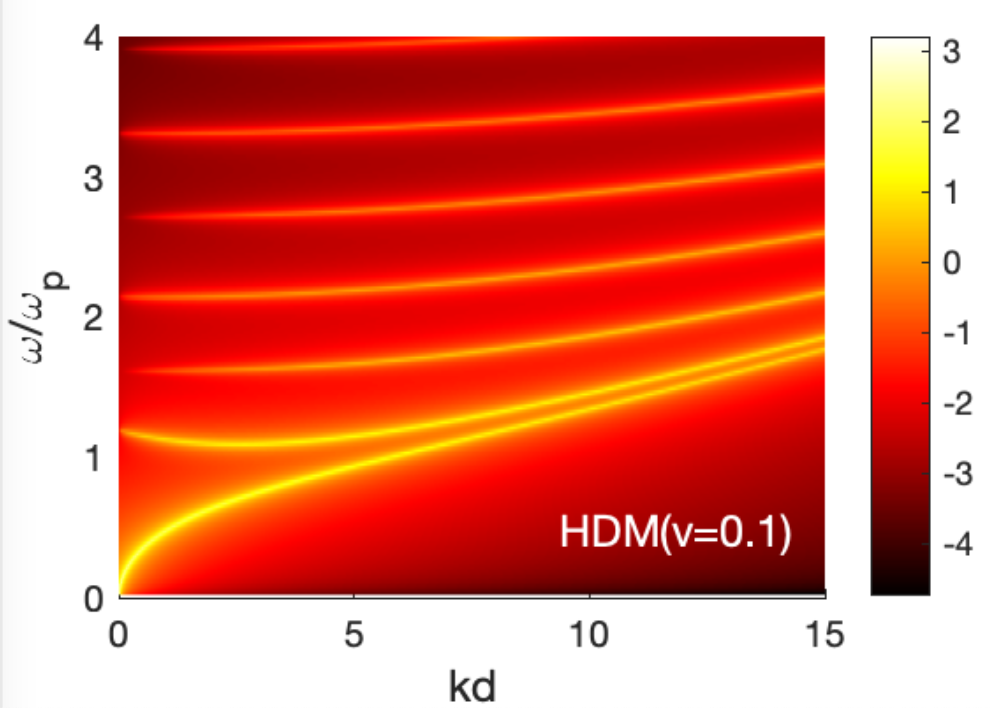}
\end{center}
\caption{Same as Fig.~\ref{fig:1} (c) but for wider range of $k$, showing that higher-frequency branches are dispersive at large $k$. $L=0.5d$ and $v$ is in unit of $d\omega_p$.\label{fig:1e}}
\end{figure} 

If $kv>\omega_p$, $Q$ will depend on $k$ via $\omega_p$ replaced by $\sqrt{\omega^2_p + k^2v^2}$. Then the spacing also depends on $k$ and the branches become dispersive with frequencies varying with $k$. This is demonstrated in Fig.~\ref{fig:1e}.

\begin{figure}
\begin{center}
\includegraphics[width=0.45\textwidth]{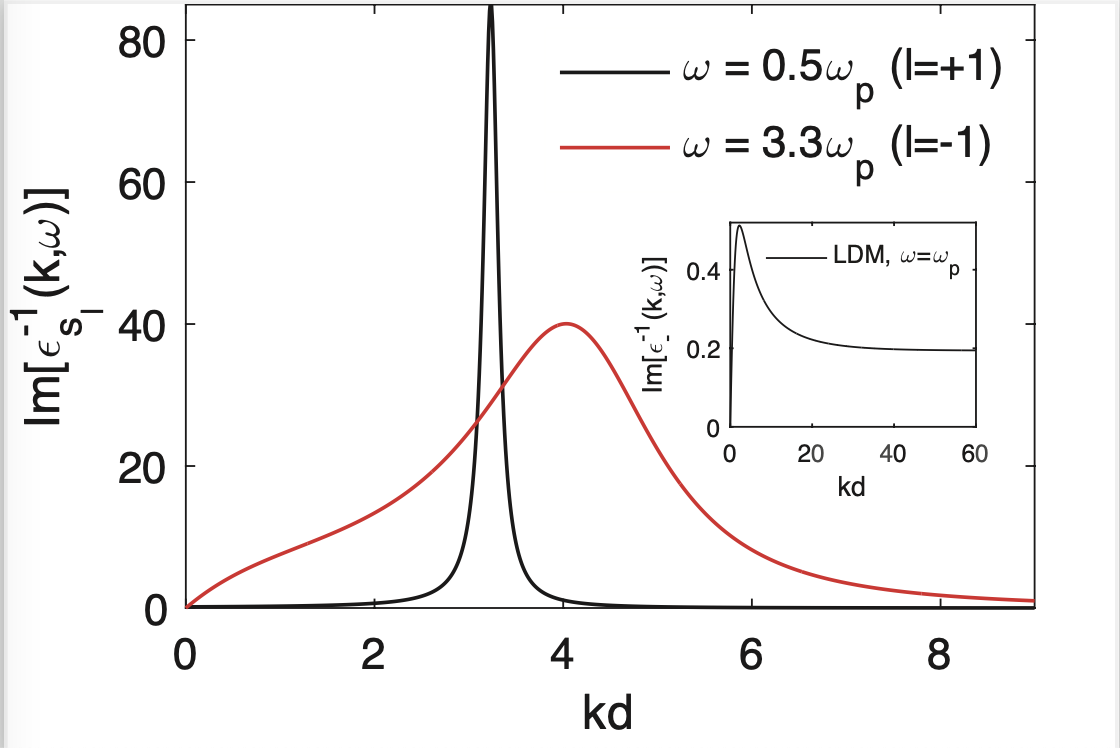}
\end{center}
\caption{Im$\left(\epsilon^{-1}_{s_l}(k,\omega)\right)$ at fixed $\omega$. For strongly dispersive branch $l=+1$, it features a sharp peak centered at $k_\omega$ satisfying $\omega = \omega_l(k_\omega,\omega)$, where $\omega_l$ is the dispersion relation of the branch $l$. For barely dispersive branch $l=-1$, the peak is broad. The red curve (for $l=-1$) has been multiplied by $20$ for visualization. Inset: Im$\left(\epsilon^{-1}_{s_l}(k,\omega)\right)$ for branch $l=-1$ of LDM, which is dispersion-less at large $k$. Parameters: $v=0.1 d\omega_p, L=0.1d$ and $\gamma = 0.05\omega_p$.\label{fig:3e}}
\end{figure} 

\begin{figure*}
\begin{center}
\includegraphics[width=0.95\textwidth]{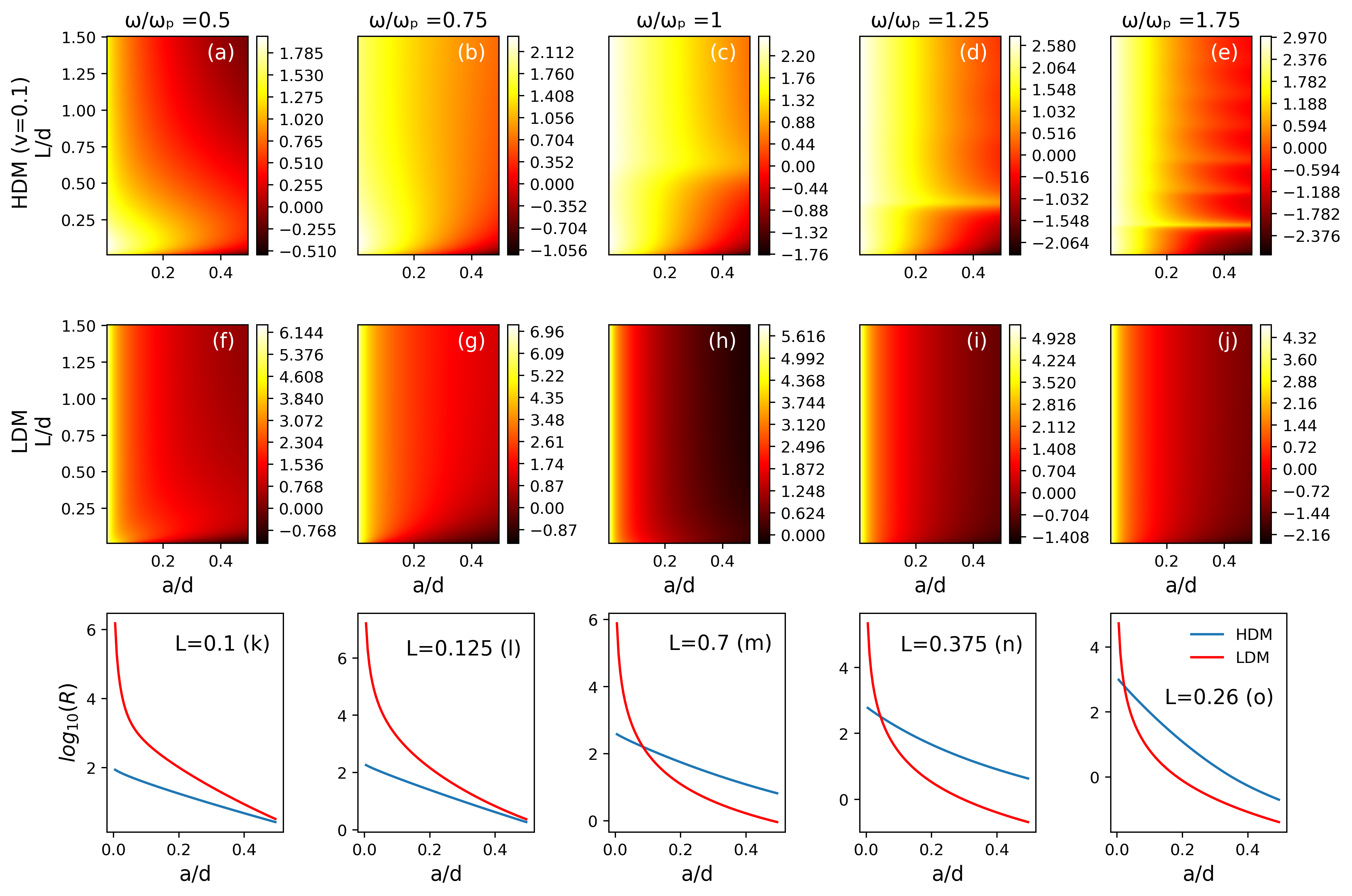}
\end{center}
\caption{$\log_{10} R$ mapped unto the $a-L$ plane for various values of $\omega$. Panels (k) - (o) are cuts at fixed $L$ as indicated, showing that $R$ diverges in LDM but remains finite in HDM for $a$ approaching zero. $\gamma = 0.05\omega_p$ and $v$ is in unit of $d\omega_p$. \label{fig:3}}
\end{figure*} 

For very large $L\gamma/v\gg1$, $\exp(-QL)$ is negligibly small due to the small but finite imaginary part of $Q$. In such case, the higher-frequency branches shall disappear. 

\begin{figure*}
\begin{center}
\includegraphics[width=0.95\textwidth]{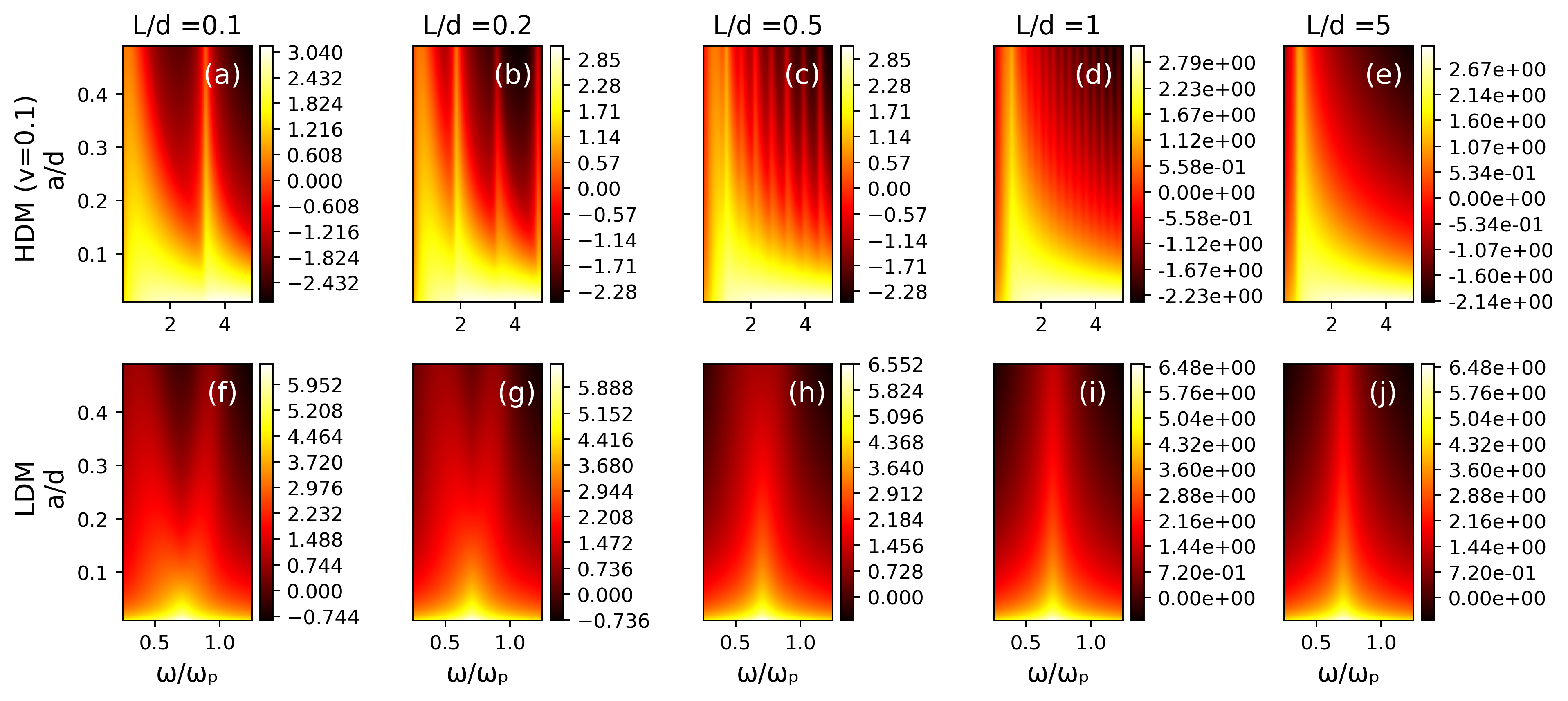}
\end{center}
\caption{$\log_{10} R$ mapped unto the $\omega-a$ plane for various values of $L$. $\gamma = 0.05\omega_p$ and $v$ is in unit of $d\omega_p$. \label{fig:4}}
\end{figure*} 

\begin{figure}
\begin{center}
\includegraphics[width=0.45\textwidth]{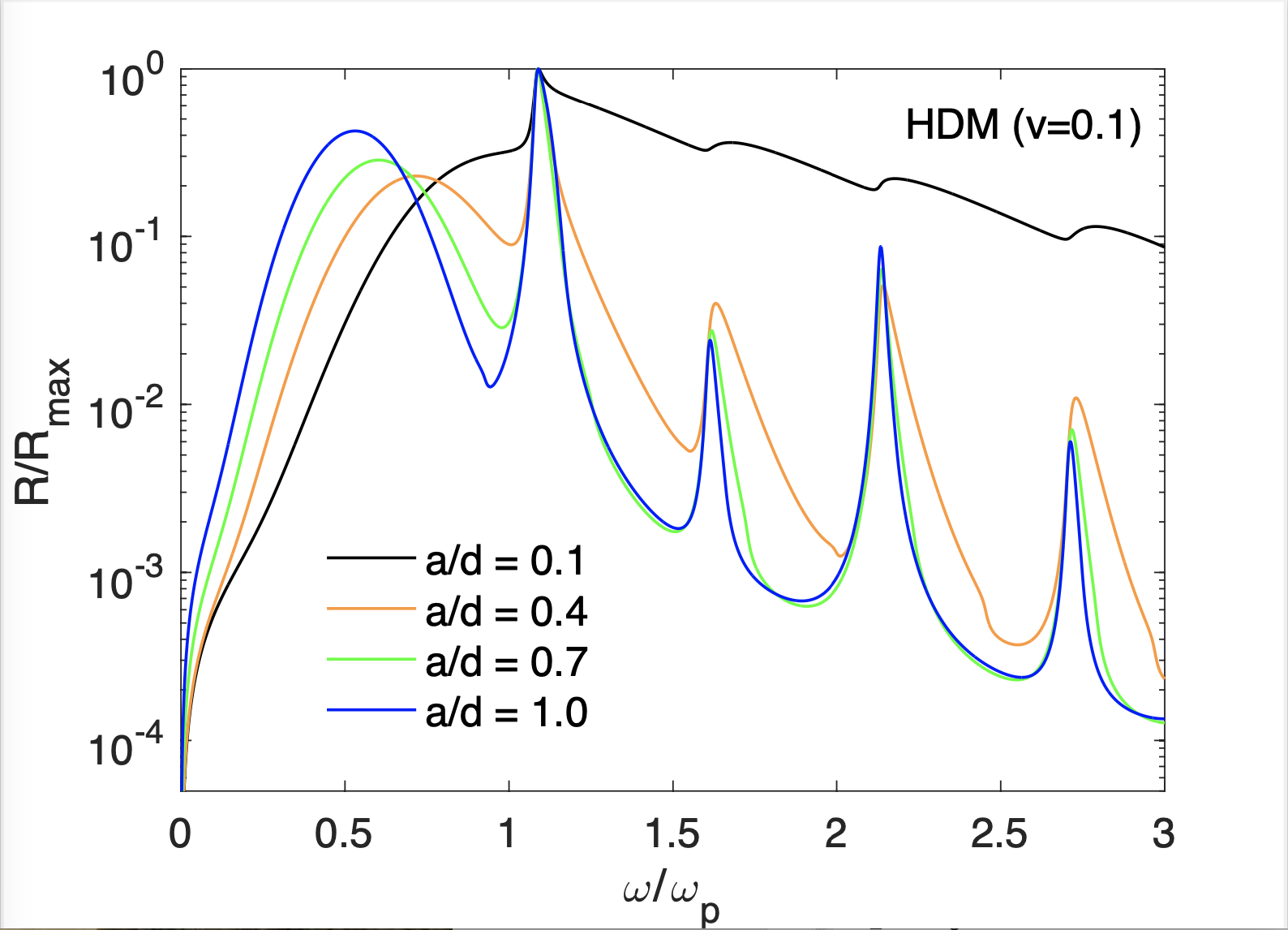}
\end{center}
\caption{Resonances with higher-frequency SPW modes revealed in the oscillatory behaviors of $R$. $R_\text{max}$ denotes the maximum of $R$ for the corresponding curve. Parameters: $\gamma = 0.01\omega_p$, $L=0.5d$ and $v$ is in unit of $d\omega_p$. \label{fig:7}}
\end{figure} 
  
\subsection{Energy transfer rate $1/\tau$}
\label{sec:3.2}
After illustrating the structure of the SPWs supported on slab surfaces, we now look into how energy is transmitted from the QD to the slab, generating SPWs in the latter. The quantity of interest is $R$ [c.f.~Eq.~(\ref{HDMR})], which directly determines the energy transfer rate $\frac{1}{\tau}$ according to Eq.~(\ref{rat}). 

\begin{figure*}
\begin{center}
\includegraphics[width=0.95\textwidth]{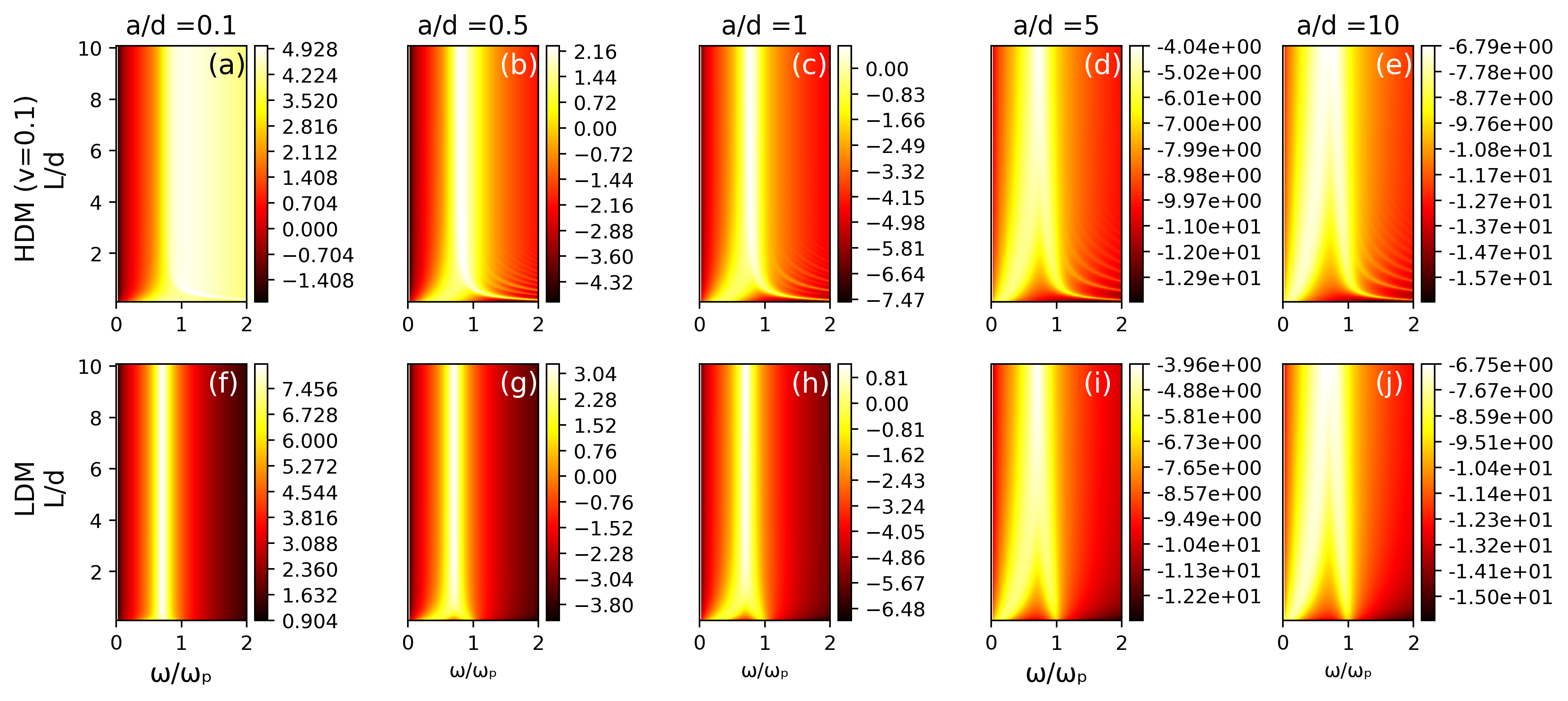}
\end{center}
\caption{$\log R$ mapped unto the $\omega-L$ plane for various values of $a$. $\gamma = 0.05\omega_p$ and $v$ is in unit of $d\omega_p$.\label{fig:5}}
\end{figure*} 

For HDM, as is clear from Fig.~\ref{fig:1}, the spectral weight $\sum_s \mbox{Im}\left(\frac{1}{\epsilon_s(k,\omega)}\right)$ is significant only for $(k,\omega)$ lying on one of the SPW branches (near or on resonance). Suppose $\omega$ cuts the $l$-th branch at $k_\omega$, i.e. $\omega = \omega_{l}(k_\omega)$, where $\omega_{l}(k)$ denotes the dispersion relation for this branch. Two situations might happen:
\begin{enumerate}[label=(\Alph*)]
\item If the branch is highly dispersive around $k_\omega$ (e.g. $l=+1$), then only the modes with $k$ near $k_\omega$ can significantly contribute: $\mbox{Im}\left(\frac{1}{\epsilon_s(k,\omega)}\right)$ displays a sharp peak centered at $k_\omega$ with a small breadth $\Delta k$ that depends on the group velocity $d\omega_l(k)/dk$ and the damping parameter $\gamma$ [Fig.~\ref{fig:3e} black curve]. From Eq.~(\ref{HDMR}), we estimate that
$R(\omega,a,L) \approx \left(k_\omega d e^{-k_\omega a}\right)^2 \left(1-s_le^{-k_\omega L}\right) W_l(\omega).$
Here $s_l$ denotes the sign of $l$ and $W_l(\omega) \approx d \Delta k \cdot \mbox{Im}\left(\frac{1}{\epsilon_{s_l}(k_\omega,\omega)}\right)$ is the total spectral weight of the peak. 
\item If the branch is barely dispersive (e.g. low-$k$ sections of branches $|l|>1$), a wide section of length $k_c$ could contribute [Fig.~\ref{fig:3e} red curve]. We may then estimate that $R(\omega,a,L) \approx  \alpha \mbox{Im}\left[\epsilon^{-1}_-(k,\omega)\right] \int^{k_c}_0 dk ~ k^2 e^{-2ka} \left(1-s_le^{-kL}\right)$, where $\alpha$ is a parameter that depends on the shape of the section in question. The integral can be explicitly evaluated but not needed here. As $L$ increases beyond $2a$, $R$ rises to a constant value $\sim (d/a)^3\left(1-e^{-2k_ca} (1+2k_ca+2k^2_ca^2)\right) \approx \frac{4k^3_cd^3}{3}e^{-2k_ca}$, where the approximation occurs for $k_ca$ not too big. 
\end{enumerate}
In both situations, $R$ rises exponentially to a constant as $a$ goes to zero, i.e. $\log R \propto a$.

In LDM, SPWs are strictly non-dispersive for $k\gg L^{-1}$ [inset of Fig.~\ref{fig:3e}]. This is similar to situation (B) but with $k_c\rightarrow\infty$. As such, $R$ diverges like $a^{-3}$ for $a$ approaching zero, as mentioned in Sec.~\ref{sec:2.3}. Note that this is so even for $\omega$ off resonance with any branches.

In Fig.~\ref{fig:3}, we display a map of $\log_{10}R$ in the $a-L$ plane at fixed $\omega$ of various values ranging from $0.5\omega_p$ to $1.75\omega_p$. It is clear -- in particular from the panels (k) - (o) -- that the above scenarios, i.e. the blow-up of $R$ in LDM and its hinderance in HDM, are vindicated. In comparison with LDM, $R$ displays a much richer structure in HDM. For LDM, regarding its dependence on $L$, $R$ simply approaches a constant $~\left(d/a\right)^{3}$ as $L$ increases [panels (f) - (j)]. For HDM, however, the $L$ dependence of $R$ is not monotonic for any $\omega$. For small $\omega$ that only resonates with the branch $l=+1$, $R$ displays a single peak at some value $L$ [panels (a) and (b)]. This can be understood from the SPW dispersion relations shown in Fig.~\ref{fig:1}. For very small $L$, the dispersion $\omega_{+1}(k)$ resonates with $\omega$ at a very large $k_\omega$, hence a very small $R$ due to the cutoff factor $e^{-2k_\omega a}$. Increasing $L$ brings $k_\omega$ to less values and larger $R$. However, as $L$ increases a crossover in the shape of $\omega_{+1}(k)$ happens: it evolves from linear to concave, resulting in smaller $\Delta k$ and a decrease in $R$ from its peak value located where the crossover takes place. On the other hand, for $\omega$ above $\omega_{-1}(k)$, $R$ displays an oscillatory structure featuring a number of peaks [panels (c) - (e)], which each indicates a resonance with one of the higher-frequency branches. Actually, as $L$ increases, these branches move down and closer and then pass through $\omega$ one by one [see Fig.~\ref{fig:1}], producing the oscillations seen in $R$. 

The higher-frequency modes are even more manifest in the $\omega-a$ map of $R$ at fixed $L$, shown in Fig.~\ref{fig:4} and Fig.~\ref{fig:7}. As seen in panels (a) - (d) here, $R$ displays very clear oscillations signifying resonances with these modes at not so small $a$. At very small $a$, though, the contributions from these modes are blurred. The happens again due to the cutoff factor $e^{-2ka}$: for very small $a$, this factor allows modes with very large $k$ to contribute, which have much closer frequencies, and damping can easily blur their individual contributions. For large $a$, only modes with small $k$ can contribute significantly, and these modes are well spaced at least for not too large $L$. If $L$ is very large, the higher-frequency modes are also closely packed and damping erases the oscillatory structure in $R$, see panel (e). In contrast, in LDM $R$ is rather pedestrian: it displays a two-branch structure at small $L$ and $a$, which is blurred otherwise for analogous reasons, see panels (f) - (j). 

Finally, in Fig.~\ref{fig:5} we display a map of $\log R$ in the $\omega - L$ plane at fixed $a$. For LDM, we see the usual SPW branches in $R$ at small $L$, which merge into a single one at large $L$ [panels (f) - (j)]. For HDM, the oscillatory structure appears on the side where $\omega>\omega_{+1}$, signifying resonances with the higher-frequency branches [panels (a) - (e)]. For large $a$, the maps for LDM and HDM look rather similar, except for the higher-frequency modes -- which are clearly visible for $L<\sim 5d$ -- and the following difference: for LDM $R$ strictly vanishes as soon as $\omega$ goes above $\omega_p$ [panel (j)], whereas for HDM it extends beyond [panel (e)]. This is so because hydrodynamic effects push up the frequencies of SPWs, as discussed in the preceding subsection. 

From Fig.~\ref{fig:5} we may also observe that, at smaller $a$ the oscillations on the high-$\omega$ side have shorter periods [panel (b)], having signals from both the symmetric and anti-symmetric higher-frequency modes; However, as $a$ increases, the signals from the symmetric modes fade out [panel (c) versus (b)], leaving only the anti-symmetric ones still discernible at fairly large $a$ [panel (e)]. At very small $a$ [panel (a)], the oscillations are not clear on the $\log R$ map, but seeable on the $R$ map, which is not shown here. 

\section{Summary}
\label{sec:4}
To summarize, we have systematically examined the nonlocal effects and size effects on the energy transfer from a dipole to a metal slab. The non-local effects are implemented using a hydrodynamic model. We have shown that the energy transfer rate, $\tau^{-1}$, can be related to the charge density-density response function of the metal slab. For HDM and LDM (as a limit of HDM), the response functions have been explicitly evaluated. The most significant finding is the existence of higher-frequency SPW modes supported on thin metal slabs. We have seen that these modes qualitatively modify the structure of $\tau^{-1}$ in the parameter space $(\omega, a, L)$. Our results demonstrate that it is imperative to build nonlocal effects into the modeling necessitated in metal-induced energy transfer imaging and other areas such as relaying information between QDs by means of metals or semiconductors~\cite{kimble2008}. 

To experimentally detect the higher-frequency SPWs, one must ensure that the slab thickness $L$ does not exceed $\sim v/\gamma$; otherwise, the hydrodynamic effects would be erased by damping. Typically, $v \sim 0.1 d\omega_p$ and hence $L$ should not be larger than $10d$ for $\gamma = 0.01\omega_p$. For most QDs, $d \sim 10~$nm, and thus $L \sim 100~$nm at the most for such parameters. Meanwhile, it is essential to have the QD frequency well tunable above the characteristic frequency of the metal, i.e. $\omega>\omega_p$ in order to reach resonance with the higher-frequency modes. Most QDs have $\omega$ amounting to a few electron volts (eVs) in energy. The slab material should then have $\omega_p$ around $1~$eV or less. Often experimented metals, such as silver and gold, have $\omega_p$ much higher and are not suitable. In addition, in those metals electrodynamic responses cannot be simply modeled by the HDM due to effects from inner atomic shells~\cite{Deng2019,liebsch1993,johnson1976}. One way out is to use doped semiconductor slabs rather than metals. For semiconductors $\omega_p$ can be tuned by doping, temperature and external field~\cite{sha1995}, and they may be the most promising candidate for experimentally studying the phenomena predicted in the present work. 

\section*{Acknowledgement}
We acknowledge the support of the Supercomputing Wales project, which is part-funded by the European Regional Development Fund via Welsh Government.

\appendix*
\section{Slab density-density response function}
\label{sec:a}
Here we derive the electrodynamic response functions for HDM. 
Assuming a time dependence $e^{-i\omega t}$, the dynamic equation (\ref{HDM}) can be recast as
\begin{equation}
-i\bar{\omega}mn_0\mathbf{V} = n_0e\mathbf{E} - mv^2 \nabla n, \quad \bar{\omega} = \omega + i\gamma.
\end{equation} 
In terms of the current density $\mathbf{J} = n_0e\mathbf{V}$ and charge density $\rho = ne$ due to the electrons, 
\begin{equation}
-i\bar{\omega} m\mathbf{J} = n_0 e^2 \mathbf{E} - mv^2 \nabla \rho. \label{J}
\end{equation}
Note that $\nabla\cdot\mathbf{E} = 4\pi (\rho + \rho_{ext})$, where $\rho_{ext}$ is the external charge density. Taking the divergence,
\begin{equation}
-i\bar{\omega} \nabla\cdot\mathbf{J} = \omega^2_p (\rho + \rho_{ext}) - v^2 \nabla^2 \rho. \label{a3}
\end{equation} 
On the other hand, we have the equation of continuity, which reads $i\bar{\omega}\rho = \nabla\cdot\mathbf{j}$ with $\mathbf{j} = \mathbf{J}\left(\theta(z)-\theta(z-L)\right)$, where $\theta(z)$ denotes the Heaviside step function: $\theta(z\leq0) = 0$ and $\theta(z>0)=1$. Combining this with Eq.~(\ref{a3}) gives
\begin{equation}
(v^2\nabla^2 + \bar{\omega}^2 - \omega^2_p) \rho = \omega^2_p \rho_{ext} - i\bar{\omega} \left(J_z(\mathbf{x}_0)\theta'(z) - J_z(\mathbf{x}_L)\theta'(z-L)\right), \label{a4}
\end{equation}
where $\mathbf{x}_{0} = (\mathbf{r},0)$ and $\mathbf{x}_L = (\mathbf{r},L)$ denote a point on the surfaces at $z = 0$ and $z=L$, respectively. From Eq.~(\ref{J}),
\begin{equation}
J_z(\mathbf{x}) = \frac{in_0e^2}{m\bar{\omega}} E_z(\mathbf{x}) + \frac{v^2}{i\bar{\omega}}\partial_z\rho(\mathbf{x}).  \label{a5}
\end{equation}
In the literature~\cite{barton1979,pendry2013}, it is often imposed that $J_z(\mathbf{x}_0) = J_z(\mathbf{x}_L) =0$. Here we do not make such imposition.  

Making use of the translation symmetry along the surfaces and assuming an overall dependence $e^{i\mathbf{k}\cdot\mathbf{r}}$ on the planar coordinates $\mathbf{r}$, that is $\rho(\mathbf{r}) = \rho(z)e^{i\mathbf{k}\cdot\mathbf{r}}$ and similarly for other field quantities, Eqs.~(\ref{4}) and (\ref{a5}) combine to yield
\begin{eqnarray}
&~&(v^2\partial^2_z + \tilde{\omega}^2 - \omega^2_p) \rho(z) = \left(\frac{\omega^2_p}{4\pi} E_z(0) - v^2 \rho'(0)\right)\theta'(z) \label{a6}\\ &~& \quad \quad \quad \quad \quad \quad \quad - \left(\frac{\omega^2_p}{4\pi}E_z(L) - v^2 \rho'(L)\right)\theta'(z-L) + \omega^2_p \rho_{ext}. \nonumber
\end{eqnarray}
Doing a cosine Fourier transform to this equation leads to
\begin{equation}
(\tilde{\omega}^2 - \omega^2_p - q^2_n v^2) \rho_n = \frac{\omega^2_p}{4\pi L_n} \left[E_z(0) - (-1)^n E_z(L)\right] + \omega^2_p \rho_{ext,n}, \label{a8}
\end{equation}
where $\rho_n$ are the cosine Fourier coefficients,
\begin{equation}
\rho(z) = \sum_n \rho_n \cos(q_nz), \quad \rho_n = L^{-1}_n \int^L_0 dz~\rho(z) \cos(q_nz). 
\end{equation}
Note that the terms involving $\rho'(0)$ or $\rho'(0)$ in Eq.~(\ref{a6}) have cancelled out and hence disappeared from Eq.~(\ref{a8}), which implies that the results do not depend on whether $J_z$ vanishes at the surfaces or not. 

One may show that
\begin{equation}
E_z(0) = - \Phi'_{ext}(0) - 2\pi \xi_0, \quad E_z(L) = - \Phi'_{ext}(L) + 2\pi \xi_L,
\end{equation}
where $\Phi_{ext}(z)$ is the external potential generated by the probe charge $\rho_{ext}$, and
\begin{eqnarray}
\xi_0 &=& \int^L_0 dz e^{-kz} \rho(z) = \sum_n \frac{k \rho_n}{k^2 + q^2_n} \left(1 - e^{-kL}(-1)^n\right), \\ \xi_L &=& \int^L_0 dz e^{k(z-L)} \rho(z) = \sum_n \frac{k \rho_n (-1)^n}{k^2 + q^2_n} \left(1 - e^{-kL}(-1)^n\right). \nonumber\\
\end{eqnarray}
Considering that $(k^2-\partial^2_z)\Phi_{ext}(z) = 4\pi \rho_{ext}$, one finds 
\begin{equation}
\frac{k^2+q^2_n}{4\pi} \Phi_{ext,n} = \rho_{ext,n} - \frac{\Phi'_{ext}(0) - (-1)^n \Phi'_{ext}(L)}{4\pi L_n}. 
\end{equation}
Using these results, Eq.~(\ref{a8}) becomes
\begin{equation}
\left(\tilde{\omega}^2 - \omega^2_p - q^2_n v^2\right) \rho_n = \frac{\omega^2_p (k^2+q^2_n)}{4\pi} \Phi_{ext,n} - \frac{\omega^2_p}{2L_n}(\xi_0 + (-1)^n \xi_L). \label{13}
\end{equation}
Note that
\begin{equation}
\xi_0 + (-1)^n \xi_L = \sum_m \frac{k \rho_m}{k^2 + q^2_m} \left(1 - e^{-kL}(-1)^m\right) \left(1+(-1)^{m+n}\right),
\end{equation}
which vanishes unless $n$ and $m$ have the same parity. This is a consequence of the reflection symmetry possessed by the slab about its mid-plane. Thus, we split $\rho(z) = \rho^+(z) + \rho^-(z)$, where $\rho^\pm(z) = \sum^\infty_{l=0} \rho^\pm_l \cos(q^\pm_lz)$ with $\rho^+_l=\rho_{2l}$, $\rho^-_l=\rho_{2l+1}$ and $q^\pm_l$ defined in the main text, and introduce
\begin{equation}
\xi_\pm = \frac{1}{2} (\xi_0 \pm \xi_L) = \sum_l \frac{k \rho^\pm_l}{k^2 + (q^\pm_l)^2} \left(1 \mp e^{-kL}\right). \label{a15}
\end{equation}
Under such provisions, Eq.~(\ref{13}) can be rewritten as
\begin{equation}
\left(\tilde{\omega}^2 - \omega^2_p - (q^\pm_l)^2 v^2\right) \rho^\pm_l = \frac{\omega^2_p (k^2+(q^\pm_l)^2)}{4\pi} \Phi^\pm_{ext,l} - \frac{\omega^2_p}{L^\pm_l} \xi_\pm. \label{a16}
\end{equation}
The above two equations can now be solved to produce 
\begin{equation}
\rho^\pm_l = \sum_{l'} \chi^\pm_{k\omega}(l,l')\Phi^\pm_{ext,l'}
\end{equation}
with the density-density response function given by
\begin{eqnarray}
&~& \chi^\pm_{k\omega}(l,l') = \frac{\omega^2_p}{\tilde{\omega}^2-\omega^2_p-v^2(q^\pm_l)^2} \label{a18} \\ &~& \quad \times \left[\frac{k^2+(q^\pm_l)^2}{4\pi} \delta_{l,l'} - \frac{1}{\epsilon_\pm} \frac{\omega^2_p}{\tilde{\omega}^2-\omega^2_p-v^2(q^\pm_{l'})^2}\frac{k(1\mp e^{-kL})}{4\pi L^\pm_{l}}\right]. \nonumber
\end{eqnarray}
For $L\rightarrow\infty$, this result reduces to what was derived in Ref.~[\onlinecite{Deng2020}] for a semi-infinite metal. 

In Eq.~\ref{a18}, we have reinstated the subscript $k\omega$, and the quantity 
\begin{equation}
\epsilon_s = 1 + \sum_l \frac{2\omega^2_s}{\tilde{\omega}^2-\omega^2_p-v^2(q^s_l)^2} \frac{k/L^s_l}{k^2+(q^s_l)^2}
\end{equation}
can be rewritten as
\begin{equation}
\epsilon_s =  1 + \frac{2k\omega^2_s}{\bar{\omega}^2 - \omega^2_p} \sum^\infty_{l=0} \frac{1}{L^s_l} \left(\frac{1}{k^2+(q^s_l)^2} - \frac{1}{Q^2 + (q^s_l)^2}\right)
\end{equation}
with $Q^2 = (\omega^2_p - \tilde{\omega}^2)/v^2$. Using the identity Eq.~(\ref{id}), we immediately get 
\begin{equation}
\epsilon_s(k,\omega) = 1 + \frac{\omega^2_{-s}}{\bar{\omega}^2 -\omega^2_p} \left(1 - \frac{k}{Q} \frac{1+se^{-QL}}{1-se^{-QL}} \frac{1-se^{-kL}}{1+se^{-kL}} \right),
\end{equation}
which after algebra gives Eq.~(\ref{eps}). Using Eq.~(\ref{a18}) in Eq.~(\ref{P}), one is led to Eq.~(\ref{17}) for $P^s_{k\omega}(l)$ in the main text.

Setting $v=0$, one gets
\begin{eqnarray}
\chi^\pm_{k\omega}(l,l') &=& \frac{\omega^2_p}{\bar{\omega}^2 - \omega^2_p} \frac{k^2+(q^\pm_l)^2}{4\pi} \delta_{l,l'} \nonumber \\ &~& \quad \quad \quad - \frac{k}{2\pi L^\pm_{l}}\frac{\omega^2_p}{\bar{\omega}^2-\omega^2_p}\frac{\omega^2_{\pm 0}}{\bar{\omega}^2 - \omega^2_{\pm 0}} 
\end{eqnarray}
as the response function for LDM.

\end{document}